\documentclass[%
 reprint,
 amsmath,
 amssymb,
 superscriptaddress,
]{jfm}
\usepackage{graphicx}
\usepackage{epstopdf, epsfig}
\usepackage{dcolumn}
\usepackage{bm}
\usepackage{amsmath,amssymb}
\usepackage{color}
\usepackage{graphicx}

\shorttitle{Viscous-like forces control the impact response of shear-thickening dense suspensions}
\shortauthor{M.-A. Brassard, N. Causley, N. Krizou, J. A. Dijksman, and A. H. Clark}

\title{Viscous-like forces control the impact response of shear-thickening dense suspensions}

\author{Marc-Andre Brassard\aff{1},  Neil Causley\aff{1}, Nasser Krizou\aff{1}, Joshua A. Dijksman\aff{2},
 \and Abram. H. Clark\aff{1}
   \corresp{\email{abe.clark@nps.edu}}
}

\affiliation{\aff{1}Department of Physics, Naval Postgraduate School, Monterey, CA USA
\aff{2}Physical Chemistry and Soft Matter, Wageningen University \& Research, Wageningen, The Netherlands}

\begin{document}

\date{\today}

\maketitle

\begin{abstract}
We experimentally and theoretically study impacts into dense cornstarch and water suspensions. We vary impact speed as well as intruder size, shape, and mass, and we characterize the resulting dynamics using high-speed video and an onboard accelerometer. We numerically solve previously proposed models, most notably the added-mass model as well as a class of {viscous-like} models. In the {viscous-like models}, the intruder dynamics are dominated by {large, viscous-like forces} at the boundary of the jammed front {where large shear rates and accompanying large viscosities are present.} We find that our experimental data are consistent with this class of models and inconsistent with the added mass model. Our results strongly suggest that the added-mass model, which is the dominant model for understanding the dynamics of impact into shear-thickening dense suspensions, should be updated to include these viscous-like forces.
\end{abstract}

\begin{keywords}

\end{keywords}



\section{Introduction}
\label{sec:intro}

A dense suspension consists of solid particles, with sizes on the scale of 1-100 $\mu$m, placed into a Newtonian fluid such that, {in the absence of external forcing or driving, the particle phase is crowded but not jammed~\citep{ohern2003,van2009jamming,Torquato2010RevModPhys}}. Such systems are common in a variety of engineering and geophysical contexts. {If a dense suspension is sheared or compressed, there can be a dramatic increase in the viscosity called shear thickening or, if the viscosity changes rapidly enough, discontinuous shear thickening (DST). A growing consensus suggests that this effect arises when short-range repulsive forces (e.g., lubrication, electrostatic, or chemical) are overcome and particles make solid-solid contact with each other~\citep{seto2013discontinuous,WyartCates2014PRL,guazzelli_pouliquen_2018}. This picture helps explain why the} rheology of the suspension varies strongly with the volume fraction $\phi$ occupied by the particles. For small particle volume fraction $\phi$ (typically $\phi<0.4$), the suspension behaves as a Newtonian fluid, with a constant viscosity $\eta$ that increases with $\phi$. For $\phi > \phi_J$ (typically $\phi_J \approx 0.6$), the particles are jammed, and the material behaves as a yield-stress solid~\citep{Brown_2014}. In between these two limits (typically $0.4<\phi<0.6$), $\eta$ increases dramatically if the shear rate $\dot{\gamma}$ exceeds some critical shear rate $\dot{\gamma}_c$, the value of which depends on $\phi$~\citep{Hoffman1972,Barnes1989,brown2009,Fall2010,Brown_2014} and other microscopic features.



Impact into a shear-thickening dense suspension (e.g., cornstarch mixed with water or another liquid at $\phi \approx 0.45-0.5$) by a foreign intruder can be similarly dramatic~\citep{lee2003ballistic,waitukaitis2012impact,peters2014quasi,han2016high,Mukhopadyay2018,han2019}. Yet, a simple application of steady-state rheology cannot explain the impact response, as the stresses predicted by DST are far too small to, e.g., support a person running across a cornstarch-water suspension; see the Introduction of Ref.~\citep{Mukhopadyay2018} for a complete discussion. {This discrepancy is not entirely surprising, since impact is a highly inhomogeneous, transient process involving both compression and shear. Experiments have repeatedly shown that the impact leads to a dynamically jammed region that grows rapidly away from the point of impact~\citep{waitukaitis2012impact,peters2014quasi,han2016high,Mukhopadyay2018}, which is thought to dominate the decelerating forces on the impacting object. The dynamically jammed region was originally thought to be associated with a local increase in $\phi$~\citep{waitukaitis2012impact,peters2014quasi}, but later work demonstrated that ``solidification proceeds without a detectable increase in packing fraction''~\citep{han2016high}. Thus, the underlying physical mechanisms that control the impact-induced solidification are still debated.} 

Regardless of the microscopic origins of the dynamic jamming during impact, the dominant theory assumes that the intruder deceleration is dominated by momentum conservation due to the growing ``added mass'' of the solidified region~\citep{waitukaitis2012impact}. {This model essentially divides the suspension into two regions: the solid-like region, which co-moves with the intruder and grows in time, and the rest of the suspension, which does not move. A third region, specifically the boundary layer between these two regions, is explicitly neglected by the added-mass model. \citet{han2016high} experimentally demonstrated the existence of a thin layer surrounding the moving, dynamically jammed region, with characteristic thickness on the order of millimeters. This thin layer separates the (moving) jammed region from the rest of the (static) suspension, so a very large shear rate $\dot{\gamma}$ must be present. Thus, this layer can be characterized by having a large, nearly constant viscosity, corresponding to what is commonly observed in the shear-thickened regime (i.e., at large shear rates) in steady-state rheology studies of dense suspensions~\citep{brown2009,fall2012shear}.} We note that added mass alone was sufficient to explain experimental data in \citet{peters2014quasi}, but these experiments were two-dimensional (2D), meaning that large, viscous-like forces would only act over a thin, quasi-1D boundary. In 3D, the relative role of viscous-like drag acting on a surface and inertial effects from changing volumes is significantly different. We return to this discussion in our conclusions, Sec.~\ref{sec:conclusion}.

Here we show via theoretical analysis and impact experiments that large, viscous-like forces at the boundary of the growing jammed mass likely play a dominant role in the dynamics of the intruder. Drawing from previous work, we theoretically analyze the case of an intruder impact into a suspension, where the dynamics include added-mass forces as well as these large, viscous-like forces at the boundary of the jammed region. We find that the original added-mass model as well as modified versions robustly predict that the maximum force $F_{\rm max}$ achieved during impact and associated time scale $t_{\rm max}$ scale with the impact velocity $v_0$ as $F_{\rm max} \propto {v_0}^2$ and $t_{\rm max} \propto {v_0}^{-1}$. These predictions are inconsistent with the data from our experiments as well as those from~\citet{waitukaitis2012impact}. These data show $F_{\rm max} \propto {v_0}^{\alpha}$ and $t_{\rm max} \propto {v_0}^{\beta}$, but with $\alpha \approx 1.5$ and $\beta \approx -0.5$ (instead of $2$ and $-1$, respectively). We find that we can better predict these observed scaling laws by assuming that viscous-like forces at the boundary of the dynamically jammed region are dominant. In addition, we consider how $F_{\rm max}$ and $t_{\rm max}$ depend on intruder size, mass, and shape, and we again find that models dominated by viscous forces at the boundary perform better than models based on added-mass. Our results suggest that the added-mass model is incomplete and may be improved by including large, viscous-like forces at the boundary of the dynamically jammed region.

\section{Theoretical Analysis}
\label{sec:theory}

Prior experiments have repeatedly demonstrated that impact into a shear-thickening dense suspension results in a dynamically jammed, solid-like region. {The boundary separating the jammed region from the rest of the suspension propagates away from the impact point at a characteristic speed that increases with packing fraction. Several theories have been proposed to explain this phenomenon. The original picture from~\citet{waitukaitis2012impact,Waitukaitis_2013} posits that the formation and dynamics of the propagating front are primarily related to volume conservation, where volume swept out by the intruder must be compensated for by compaction of particle phase. However, as mentioned above, subsequent work has demonstrated that shear jamming is a more likely cause~\citep{han2016high,Han2018shearfronts,han2019PRL}. Thus, Reynolds dilatancy, which is closely related to shear-jamming~\citep{bi2011jamming,Wang2018}, likely plays a crucial role in the front dynamics. Darcy-Reynolds theory, experimentally tested by \citet{jerome2016}, also demonstrated that there is a strong coupling between dilation and fluid flow through the pore structure. Thus, for very small particles like cornstarch, the Darcy pressure is proportional to the inverse of particle size squared. This may explain why the growing jammed front remains quasi-solid and relaxes slowly; we return to this point in our discussion of the force relaxation dynamics in Sec. ~\ref{sec:relax}. Our perspective in this paper is to assume that the growing, dynamically jammed region exists and to remain agnostic as much as possible regarding its microscopic origins.

\citet{waitukaitis2012impact} also showed that the impact response was not primarily due to the dynamically jammed region reaching the boundary of the system and thus connecting the intruder to the boundary. They verified this by changing the depth of the suspension and demonstrating that the impact response did not depend on it (we also verify this for our experiments, as described below). When the dynamically jammed region reaches the boundary, there is a very large increase in the force felt by the intruding object, which has also been seen in other studies~\citep{Maharjan2018,peters2014quasi}. Here, we focus on the regime where system-spanning dynamically jammed regions do not occur.}

\subsection{Quasi-one-dimensional front development}

If the front propagation process is assumed to be quasi-one-dimensional, meaning the compacted region grows only downward in a column with depth $z_f$ and not laterally, then the volume-conservation picture~\citet{Waitukaitis_2013} can be applied to obtain 
\begin{equation}
    k \equiv \frac{v_f}{v} = \frac{\phi_J}{\phi_J - \phi_0}.
\end{equation}
where $v_f = dz_f/dt$ is the characteristic speed of the front and $v = dz/dt$ is the intruder's speed. For a schematic, see Fig.~\ref{fig:front-growth}a. {Despite the fact that no detectable compaction of the suspension occurs,} this dependence on $\phi_0$ and $\phi_J$ was corroborated by~\citet{peters2014quasi} and \citet{han2016high} for 2D and 3D impacts, where $\phi_J\approx 0.51$ is the jamming packing fraction for the cornstarch particles, low due to swelling~\citep{chen2019discontinuous}. This implies $k \approx 10$ when $\phi_0 = 0.46$. {Since we do not vary $\phi$ in our experiments, we will assume a constant value of $k$.}


\subsection{Including lateral front growth} \label{sec:lateral-growth}
Experiments show that the jammed region below the intruder does not grow strictly downward in a quasi-1D column but spreads out laterally as well, albeit at a smaller speed. Experiments with small impacting objects~\citep{peters2014quasi,han2016high} typically find that the transverse dimension is about half of $z_f$, meaning that the volume of the jammed region still scales as ${z_f}^d$, where $d$ is the dimensionality of the system (2 or 3). Additionally, experimental data show that the front slows down as it moves; see Fig. 4 of~\citet{peters2014quasi} and Fig. 4 of ~\citet{han2016high}. The data from these papers appears consistent with $z_f \propto z^\Gamma$ with $\Gamma<1$. {We discuss models below including this slowing-down behavior, corresponding to $z_f \propto z^\Gamma$, where $\Gamma$ is varied to demonstrate that the results are relatively insensitive to its value. If compression-induced jamming were the sole cause, the volume of the jammed region scales as ${z_f}^d$ while the volume swept out by the intruder is linear $z$, meaning that $z_f$ scales as $z^{1/d}$. So, $\Gamma = 1/d$ represents a reasonable lower bound for $\Gamma$.} 

\subsection{General equation of motion}

To understand how the front growth, including its shape, affects the resulting dynamics, we consider a generic equation of motion that describes the dynamics of the intruder. Assuming that the growing solid-like region is rigidly connected to the intruder, then the total momentum of both is $p = [m+m_a(t)]v(t)$, where $m$ is the constant intruder mass, $m_a(t)$ is the added mass, and $v(t)$ is the velocity of the intruder and solid-like region. The shape of the jammed region and its growth rate will set $m_a(t)$. To complete the equation of motion, there are three external forces to consider. Two relate to gravity: the weight of the intruder, $F_g = mg$, and a buoyant force $F_b$ from the displaced suspension (this term is typically negligible). The third, which is not included in the added-mass model, is any viscous-like forces $F_v$ that act at the boundary of the jammed region. Newton's second law can then be written as
\begin{equation}
    (m+m_a)\frac{dv}{dt} + v\frac{dm_a}{dt} = F_b + F_g + F_v.
    \label{eqn:F=ma}
\end{equation}

\begin{figure}
    \centering
   \includegraphics[width=0.9\columnwidth]{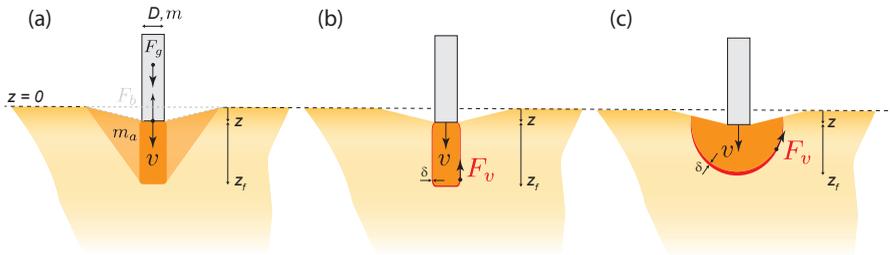}
    \caption{Three cases for front growth and ensuing intruder dynamics: (a) the original added mass model with downward and lateral growth, (b) a viscous model with only downward front growth, and (c) a front that grows downward and laterally and can experience added-mass and viscous forces. { In all three panels, the dashed line indicates the initial surface at $z=0$, $z$ indicates the depth below the initial surface and $z_f$ indicates the length scale of the jammed region.}}
    \label{fig:front-growth}
\end{figure}

Before Eq.~\eqref{eqn:F=ma} can be solved for $z(t)$ (including the derivatives $v(t) = dz/dt$, and the acceleration $a(t) = dv/dt$), assumptions must be made about the mathematical form of $m_a$, $F_b$, and $F_v$. Based on our front dynamics discussion, we now consider a few scenarios, shown in Fig.~\ref{fig:front-growth}, and solve Eq.~\eqref{eqn:F=ma} numerically or, if possible, exactly. The original added mass model, shown in panel (a), assumed that a solid, cylindrical plug grows straight down, but that the total added mass is some proportion of an inverted cone-shaped region that grows downward and outward at the same rate. {We note that schematic in Fig.~\ref{fig:front-growth}(a), which follows~\citet{waitukaitis2012impact}, does not show a cone but a truncated cone; this difference does not affect the scaling laws shown}. An alternative, shown in panel (b) is to consider drag force $F_v$ on the growing solid plug through shear in a boundary layer with thickness $\delta$. A version of this model was proposed in Appendix D of \citet{waitukaitis2014impact}. {Although the existence of this thin boundary layer was documented in~\citet{han2016high}, the microscopic physics that set $\delta$ are not known. Some possibilities include that $\delta$ is set by $\phi$, similar to the front width in~\citep{Waitukaitis_2013}, or by the particle size, perhaps through Darcy pressure.} Finally, several 2D and 3D imaging experiments suggest that the solid-like region grows laterally. In this case, the solid region experiences a drag force that grows with its surface area, as depicted in panel (c). We first describe the first two cases in Sections~\ref{sec:addedmassmodel} and \ref{sec:viscousmodel}. We then discuss scaling laws predicted by these models in Sections~\ref{sec:scaling-addmass} and \ref{sec:scaling-visc}. Finally, we discuss the third case and its scaling laws in Section~\ref{sec:hybrid-scaling}.

\subsection{Case 1: added-mass model, no viscous drag \label{sec:addedmassmodel}} 
First, we consider the original added-mass model~\citep{waitukaitis2012impact}, which assumed that the solidified region is an inverted cone whose height and radius grow at the same rate $v_f = k v$, yielding $m_a=C_m \rho (1/3)\pi(D/2+kz)^2kz$, where $\rho$ is the suspension mass density and $C_m$ is an added mass coefficient found experimentally to be $C_m\approx 0.37$. The fact that $C_m<1$ means that the entirety of the added mass region is not perfectly rigidly connected to the intruder. They assumed that viscous drag was negligible or absent and set $F_v = 0$ and that $F_b$ comes from displaced fluid in a conical depression near the intruder, $F_b = 1/3\pi\rho g z(D/2+kz)^2$. Numerical solutions to this model are qualitatively similar to experimental trajectories, as shown in Fig.~\ref{fig:models-and-expt} (thick black dashed line) and in~\citet{waitukaitis2012impact}.

\subsection{Case 2: cylindrical jammed region with large, viscous-like drag \label{sec:viscousmodel}}
\begin{figure}
    \centering
    \raggedright (a) \hspace{60mm} (b) \\ \centering
    \includegraphics[trim = 0mm 0mm 0mm 0mm, clip, width = 0.49\columnwidth]{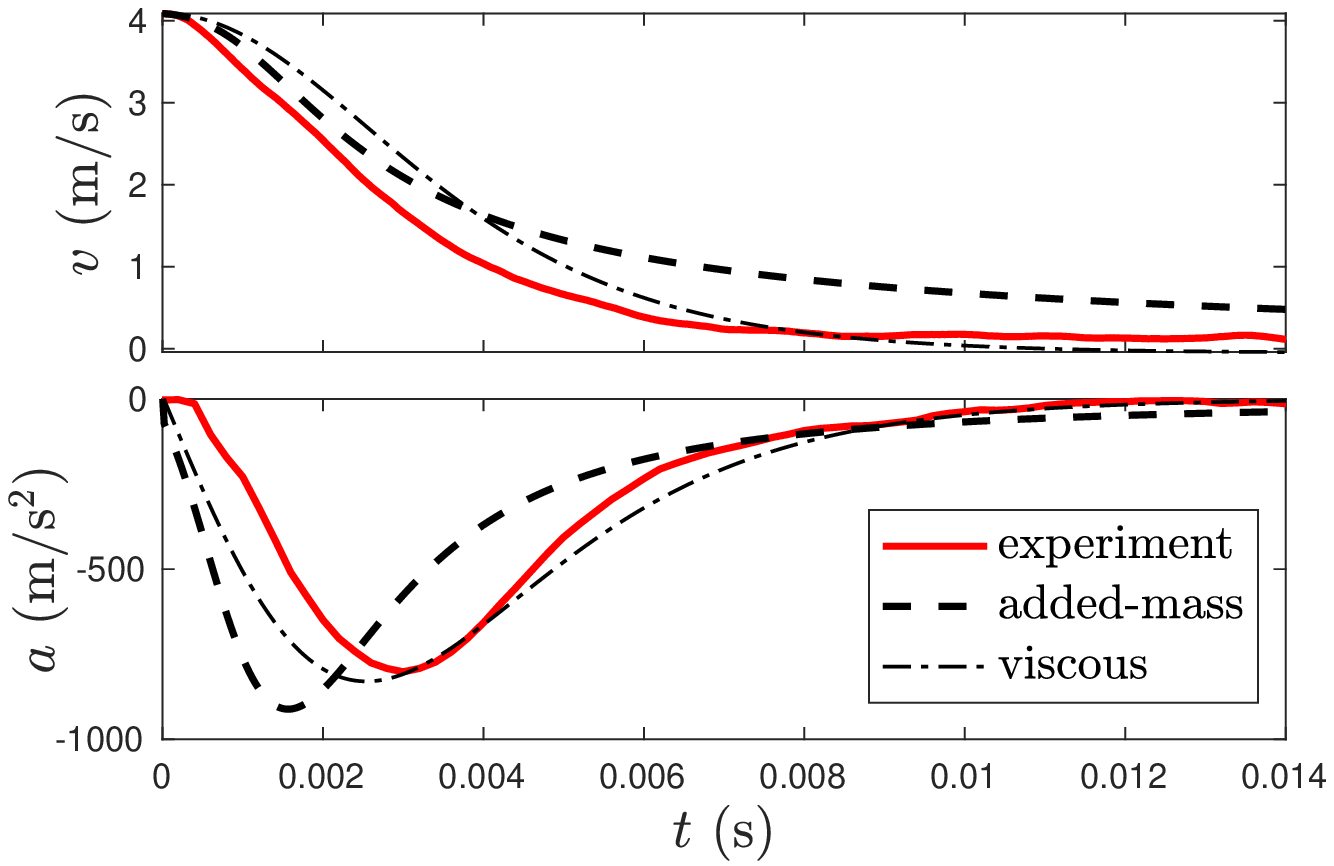}
    \includegraphics[trim = 0mm 0mm 0mm 0mm, clip,width = 0.49\columnwidth]{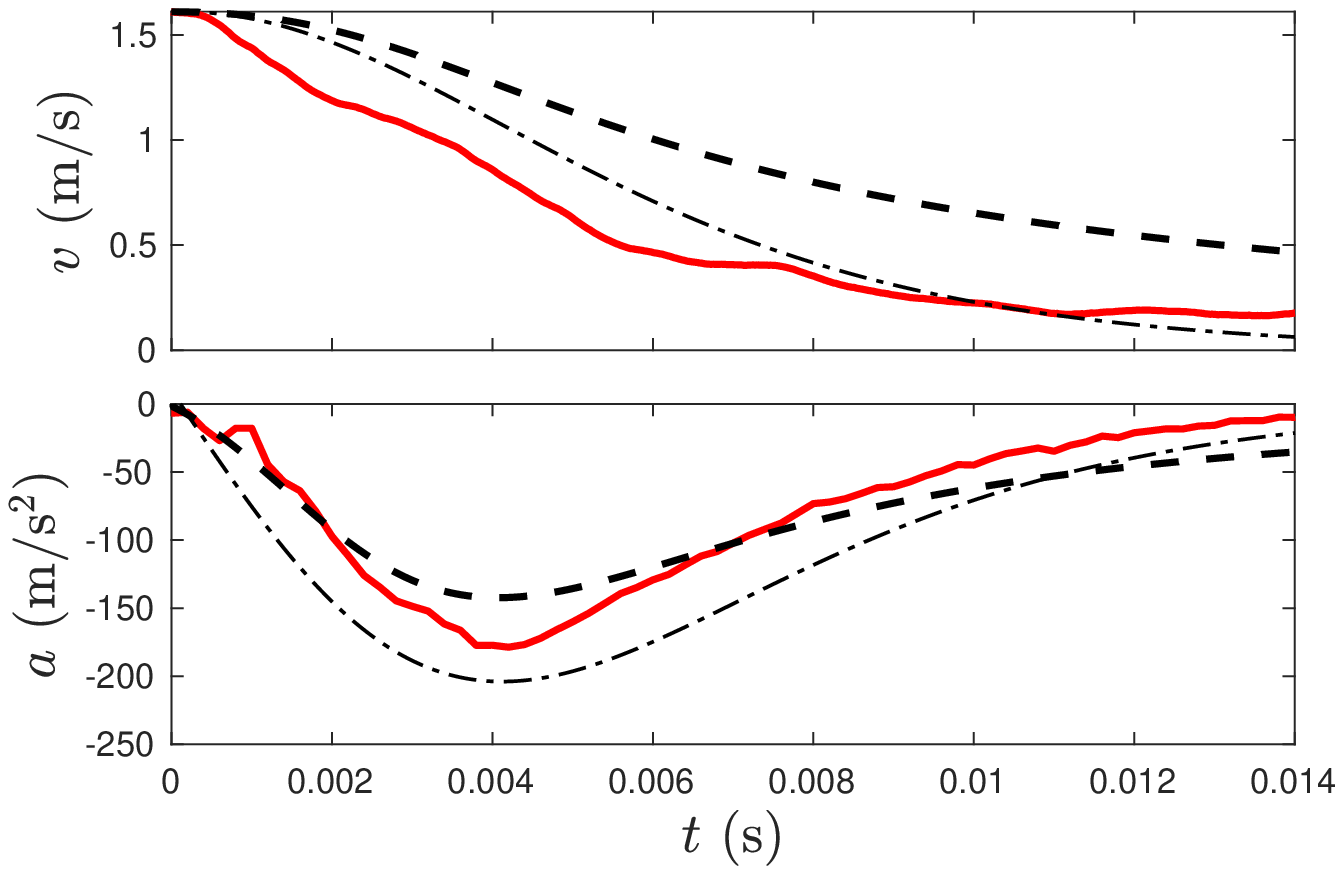}
    \caption{Experimental trajectories (red) for velocity $v(t)$ and acceleration $a(t)$ are compared with the solutions to the intruder equation of motions for the added-mass (thick black dashed line) and viscous (thinner black dot-dashed line) models discussed in Sections~\ref{sec:addedmassmodel} and \ref{sec:viscousmodel}, respectively. Panels (a) and (b) show impacts with $v_0 \approx 4$~m/s and $v_0 \approx 1$ using cylindrical intruders with $m=189$~g and $D=25.4$~mm. The fit parameters in both models can be adjusted to approximately match the experiments. The added-mass model trajectories for both panels were solved using $C_m = 0.1$, which is much smaller than the value $C_m = 0.37$ used in \citet{waitukaitis2012impact}. The viscous model trajectories are those in Eqs.~\eqref{eqn:v-viscous} and \eqref{eqn:a-viscous} using $C_v = 0.4$, $\eta_s = 20$~Pa$\cdot$, and $\delta = 1$~mm.}
    \label{fig:models-and-expt}
\end{figure}

While the added-mass model provides qualitative features that can be matched to experiments, it is not unique in this respect: other choices for $F_v$ and $m_a$ yield similar results and can also be calibrated to match experimental trajectories, as also shown in Fig.~\ref{fig:models-and-expt} (thinner black dot-dashed line). In particular, models where $F_v$ is dominant yield similar results and have the advantage of matching other features of the dynamics, as discussed below. As shown in Fig. 2 of~\citet{han2016high}, the growing jammed region moves at approximately the same speed $v$ as the intruder, and it is surrounded by a thin layer of thickness $\delta \approx 5$~mm where the shear rate is $v/\delta$. Thus, on dimensional grounds, we can approximate viscous force as
\begin{equation}
    F_v=-C_v \eta_s S \frac{v}{\delta}.
    \label{eqn:drag_force}
\end{equation}
Here  $C_v$ is a dimensionless drag coefficient, $\eta_s$ is the effective (constant) viscosity of the suspension, and $S$ is the surface area of the jammed region. {We reemphasize that our use of a constant viscosity $\eta_s$ does not mean that we consider the suspension to be a simple, viscous fluid. Instead, we refer to the large, nearly constant viscosity observed for $\dot{\gamma}>\dot{\gamma}_c$. For example, Fig. 11 of~\citet{fall2012shear} shows a nearly constant value of $\eta_s \approx 100$~Pa$\cdot$s for cornstarch suspensions in the fully shear-thickened regime for several different volume fractions. Additionally, the shear rates in this layer would be on the order of $\dot{\gamma}\sim 10-100$~s$^{-1}$, which is larger than $\dot{\gamma}_c \sim 1-10$ for cornstarch suspensions at volume fractions near $\phi = 0.4$~\citep{fall2012shear}.}

This model is a generalized form of a model appearing in Appendix D of Waitukaitis' Ph.D. thesis~\citep{waitukaitis2014impact}, which involves a columnar, solid-like front growing beneath the intruder with height $h_f = k z$ and thus $S = \pi D k z + \pi D^2 / 4$. We note that their dimensional analysis used $D$ in place of $S/\delta$, which then required $\eta_s$ to be unphysically large, $\eta_s \approx 2000$. This situation is sketched in Fig~\ref{fig:front-growth}(b). If the second term is dropped on the grounds that $D$ is much smaller than $kz$ or that the viscous-like forces only act on the sides of the growing cylinder, then the resulting dynamics are exactly solvable. ({This approximation is certainly not valid during the very early stages of impact, which would be particularly important for large $D$ and small $m$}.) In this case, Eq.~\eqref{eqn:drag_force} becomes
\begin{equation}
    F_v=-C_v \pi D \eta_s k z \frac{v}{\delta}.
    \label{eqn:drag_force2}
\end{equation}
Assuming other forces can be neglected and $m_a$ is negligible, then Eq.~\eqref{eqn:F=ma} can be exactly solved, yielding: 
\begin{align}
    v(t)&=v_0 {\rm sech}^2(t/\tau),
    \label{eqn:v-viscous} \\
    a(t)&=-\sqrt{\frac{2C_v \pi D k \eta_s}{m\delta}}{v_0}^{3/2}{\rm sech}^2(t/\tau){\rm tanh}(t/\tau),
    \label{eqn:a-viscous} 
\end{align}
where $\tau = \sqrt{C_v \pi D k \eta_s v_0/2m\delta}$. These functions are plotted in Fig.~\ref{fig:models-and-expt} and agree well with experiments. The viscous model solution in Fig.~\ref{fig:models-and-expt} use $C_v = 0.42$, $\delta = 0.5$~cm, and $\eta_s = 100$~Pa$\cdot$s. This agrees well with the viscosity of cornstarch and water-CsCl suspensions in the shear-thickening regime with similar values of $\phi$, as shown in Fig.~11 of~\citet{fall2012shear}. This comparison demonstrates that reasonable parameter values can be used in matching to experiments, although there is some flexibility and therefore uncertainty in the values of these parameters. 

\subsection{Scaling laws for the added mass model}
\label{sec:scaling-addmass}
Since both the added mass and the viscous drag model can be reasonably matched to experimentally observed intruder trajectories, some further validation can come from comparing how $F_{\rm max}$ and $t_{\rm max}$ scale with $v_0$, $m$, and $D$. Such scalings have been previously used in the case of impact to connect macroscale dynamics with the microscale mechanisms that give rise to them~\citep{Walsh2003,Uehara2003,goldman2008,clark2014,zhao2015granular,Krizou2020}. For all experiments and theoretical models, we find that these scalings can be well approximated by
\begin{align}
    F_{\rm max} &= A {v_0}^\alpha \label{eq:F_max_A} \\
    t_{\rm max} &= B {v_0}^\beta \label{eq:t_max_B}.
\end{align}
The prefactors $A$ and $B$ can vary with intruder properties, and we will examine how they depend on $m$, $D$, and, for conical intruders, cone angle $\theta$.

{In the added-mass model, $F_{\rm max}$ and $t_{\rm max}$ can depend on four parameters: $m$, $\rho$, $v_0$, $D$. A total of three dimensionless quantities can be formed by these six total parameters, so $F_{\max}/\rho{v_0}^2 D^2$ and $t_{\max} v_0/D$ must each be a function of a single dimensionless quantity $m/\rho D^3$. This suggests that, when $m/\rho D^3$ is fixed, $F_{\rm max}\propto {v_0}^2$ and $t_{\rm max}\propto {v_0}^{-1}$.} The added-mass model was solved numerically by~\citet{Mukhopadyay2018}, finding that $F_{\rm max} \propto {v_0}^2 m^{2/3}$. We also numerically solve the added-mass model and find the same result, along with $t_{\rm max} \propto {v_0}^{-1}m^{1/3}$. We find $F_{\rm max}$ and $t_{\rm max}$ to be nearly independent of $D$ in the range of parameters studied here, in agreement with~\citet{Mukhopadyay2018}. {The lack of dependence on $D$, combined with the dimensional analysis, also requires $F_{\max} \propto \rho{v_0}^2 D^2 (m/\rho D^3)^{2/3}$ and $t_{\max} \propto D/{v_0} (m/\rho D^3)^{1/3}$, in agreement with the numerical solutions.}


\begin{figure}
    \centering
    \raggedright \hspace{15mm}(a) \hspace{45mm} (b) \\ \centering
    \includegraphics[trim = 0mm 0mm 0mm 0mm, clip, width = 0.39\columnwidth]{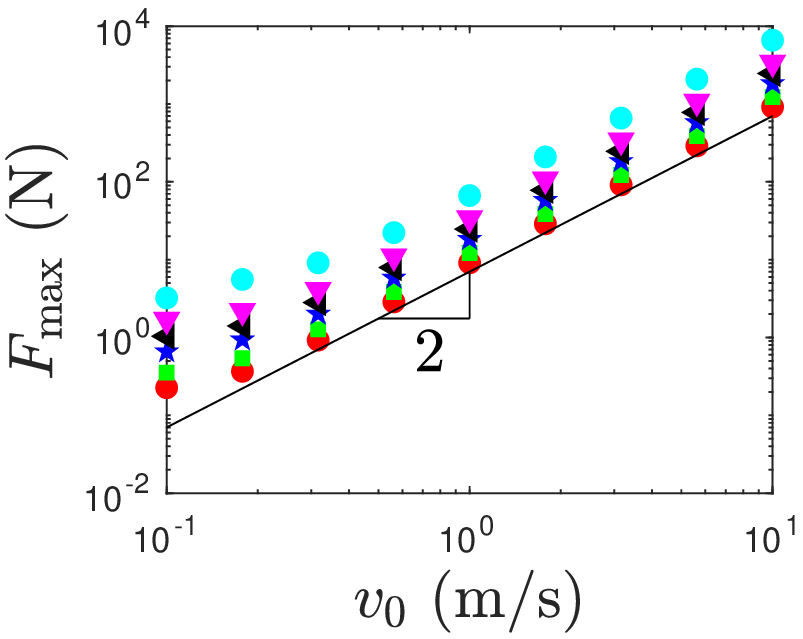}
    \includegraphics[trim = 0mm 0mm 0mm 0mm, clip, width = 0.39\columnwidth]{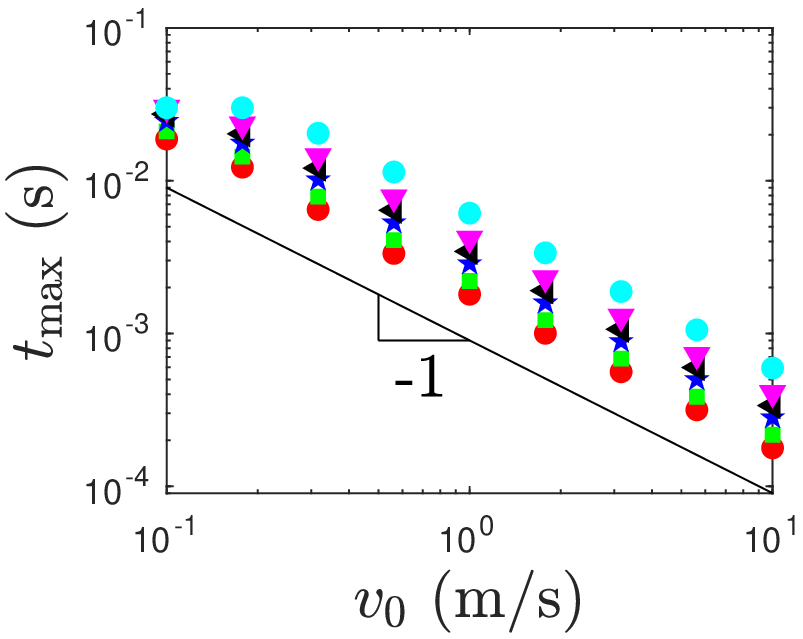} \\
    \raggedright  \hspace{15mm} (c) \hspace{45mm} (d) \\ \centering
    \includegraphics[trim = 0mm 0mm 0mm 0mm, clip, width = 0.39\columnwidth]{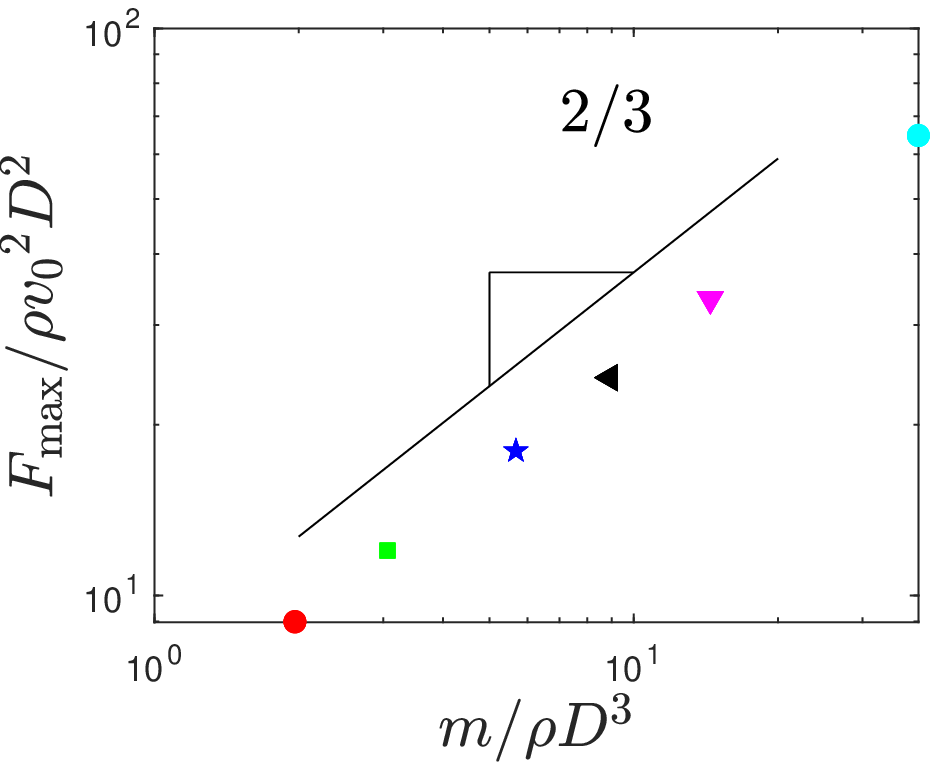}
    \includegraphics[trim = 0mm 0mm 0mm 0mm, clip, width = 0.39\columnwidth]{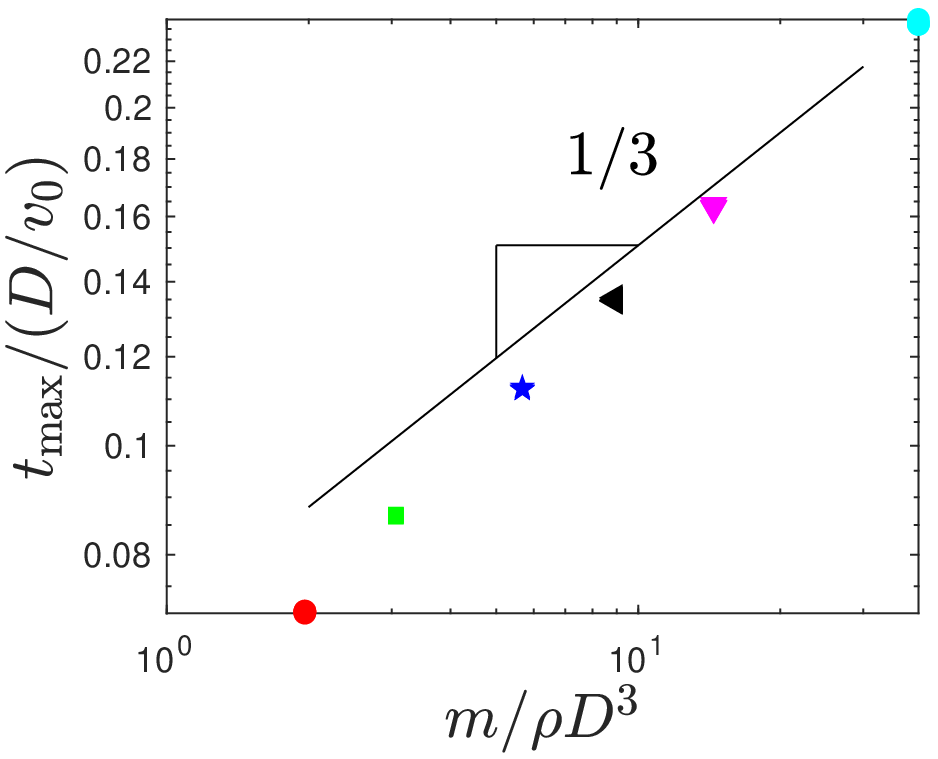}
    \caption{Panels (a) and (b) show $F_{\rm max}$ and $t_{\rm max}$ plotted as a function of $v_0$, where each symbol represents a different mass. All data with $v_0>0.5$~m/s [to avoid the low-velocity plateau seen in (a) and (b), where gravity begins to play a role] is plotted in panels (c) and (d) in terms of dimensionless quantities $F_{\rm max}/\rho {v_0}^2 D^2$, $t_{\rm max}/(D/v_0)$, and $m/\rho D^3$}
    \label{fig:added-mass-sims}
\end{figure}

\subsection{Scaling law for viscous models}
\label{sec:scaling-visc}
{ In the viscous model, there are two additional dimensionless numbers: ${\rm Re} = \rho v_0 D/\eta_s$ and $D/\delta$. This means that the dimensional analysis no longer directly predicts a relationship between $F_{\rm max}$ and $v_0$. However,} the peak force and time at which the peak acceleration occurs can be directly calculated by differentiating $a(t)$ in Eq.~\eqref{eqn:a-viscous}, setting the result to zero, and solving for $t$. This time $t_{\rm max}$ can be substituted back into Eq.~\eqref{eqn:a-viscous} to calculate $a_{\rm max} = a(t_{\rm max})$. By this method, the peak force $F_{\rm max} = m a_{\rm max}$ and $t_{\rm max}$ found to be:
\begin{align}
    F_{\rm max}&=\sqrt{2C_v D k \eta_s m/\delta}{v_0}^{3/2} {\rm sech}^2(\beta)\tanh(\beta),
    \label{eqn:a_peak_viscous} \\
    t_{\rm max}&=\sqrt{\frac{2 m \delta}{C_v D k \eta_s v_0}}\beta,
    \label{eqn:t_peak_viscous} 
\end{align}
where $\beta = \frac{1}{2}{\rm log}(2+\sqrt{3})$. We also solve the viscous model numerically and find the same result. Thus, for the viscous model where the jammed region is a cylindrical column, $F_{\rm max} \propto {v_0}^{3/2} m^{1/2}D^{1/2}$ and $t_{\rm max} \propto {v_0}^{-1/2} m^{1/2}D^{1/2}$. {These scaling relations can also be written in terms of the dimensionless quantities given above, e.g., $F_{\max}/\rho{v_0}^2 D^2 \propto {\rm Re}^{-1/2}(D/\delta )^{1/2}(m/\rho D^3)^{1/2}$.}

\subsection{Case 3: hybrid models and scaling laws}
\label{sec:hybrid-scaling}

The simple viscous model discussed above does not include several features that may make a comparison with experimental data more complicated. First, the solidified region grows in all three dimensions, not just straight down in a cylindrical column. This is particularly relevant when the contact point for the impacting object is point-like. Second, momentum is being transferred to the solidified region, so added-mass terms must be included generally. Third, the rate at which the front is growing tends to decrease with time (i.e., $\Gamma<1$). Finally, many of the physical parameters in these models have been previously studied, so there are physical bounds on, e.g., $\eta_s$ based on experimental measurements. To check how sensitive the scaling laws are for varying shapes of the growing jammed region, we study a case where the volume of the solidified region grows as ${z_f}^3$ and the surface area grows as ${z_f}^2$, where $z_f = k z^\Gamma$. For simplicity, we approximate the growing jammed region as a hemisphere where the volume and surface area are $2\pi{z_f}^3/3$ and $2\pi{z_f}^2$, respectively. This situation is sketched in Fig~\ref{fig:front-growth}(c). We note that this overestimates the volume (and thus the added mass) in particular, since ~\citet{han2016high} showed that the mass is better approximated as a half-ellipsoid with semi-minor axes of $z_f$, $z_f/2$, and $z_f/2$, meaning that the volume is $1/4$ of the hemisphere. This yields
\begin{align}
    m_a &= C_m  \rho \frac{2\pi}{3} (kz^\Gamma)^3,
    \label{eqn:m_a_spread} \\
    F_v &= - 2 C_v \eta_s \pi (kz^{\Gamma})^2 \frac{v}{\delta}.
    \label{eqn:F_v_spread}
\end{align}

We first consider the case where the added-mass term is dominant. We set $C_m=0.2$, $C_v=0$, $F_b = 0$, and numerically solve Eq.~\eqref{eqn:F=ma}. We find that $F_{\rm max} \propto {v_0}^{2}$ and $t_{\rm max} \propto {v_0}^{-1}$ persist for all values of $\Gamma$ we study ($1/3<\Gamma<1$). We emphasize that these scaling exponents are {required by the dimensional analysis} and inconsistent with experimental data shown below. Next, we consider the case where the viscous term is dominant, setting $C_v=0.5$, $C_m=0$, and $F_b=0$. For $\Gamma = 1$, corresponding to the front moving at a constant speed in all directions, we find $F_{\rm max} \propto {v_0}^{1.63}m^{0.69}$ and $t_{\rm max} \propto {v_0}^{-0.67}m^{0.33}$. For $\Gamma = 1/3$, corresponding to the lower-bound case where compression-induced jamming is dominant (see Section~\ref{sec:lateral-growth}), we find $F_{\rm max} \propto {v_0}^{1.40}m^{0.40}$ and $t_{\rm max} \propto {v_0}^{-0.40}m^{0.60}$. The behavior smoothly varies between these limits as $\Gamma$ is varied. For example, when $\Gamma = 0.7$, we find $F_{\rm max} \propto {v_0}^{1.58}m^{0.59}$ and $t_{\rm max} \propto {v_0}^{-0.60}m^{0.42}$.

\begin{figure}
    \centering
    \raggedright \hspace{15mm}(a) \hspace{45mm} (b) \\ \centering
    \includegraphics[trim = 0mm 0mm 0mm 0mm, clip, width = 0.39\columnwidth]{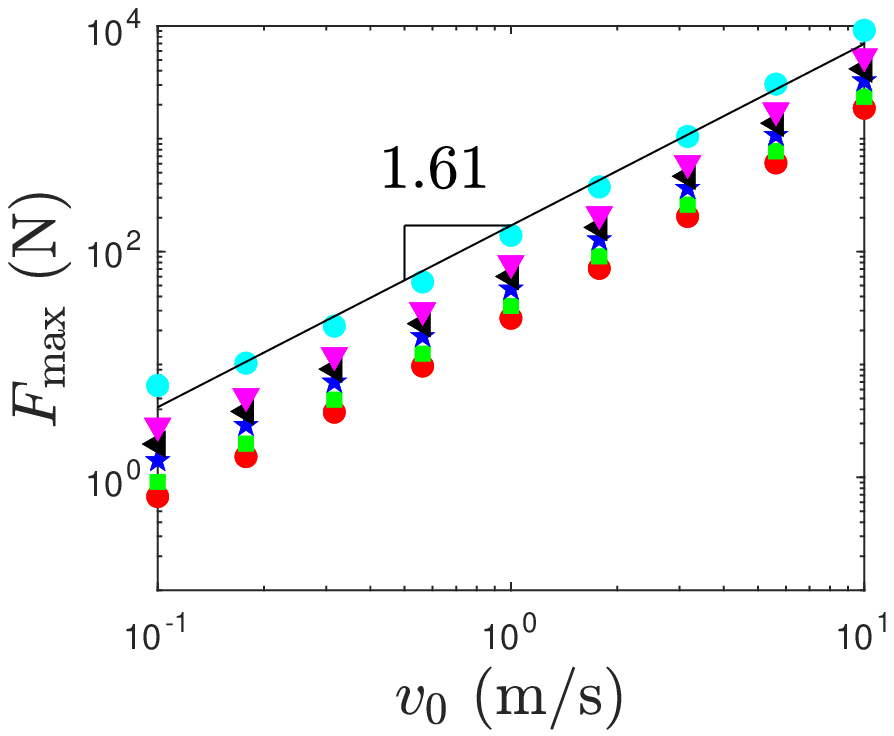}
    \includegraphics[trim = 0mm 0mm 0mm 0mm, clip, width = 0.39\columnwidth]{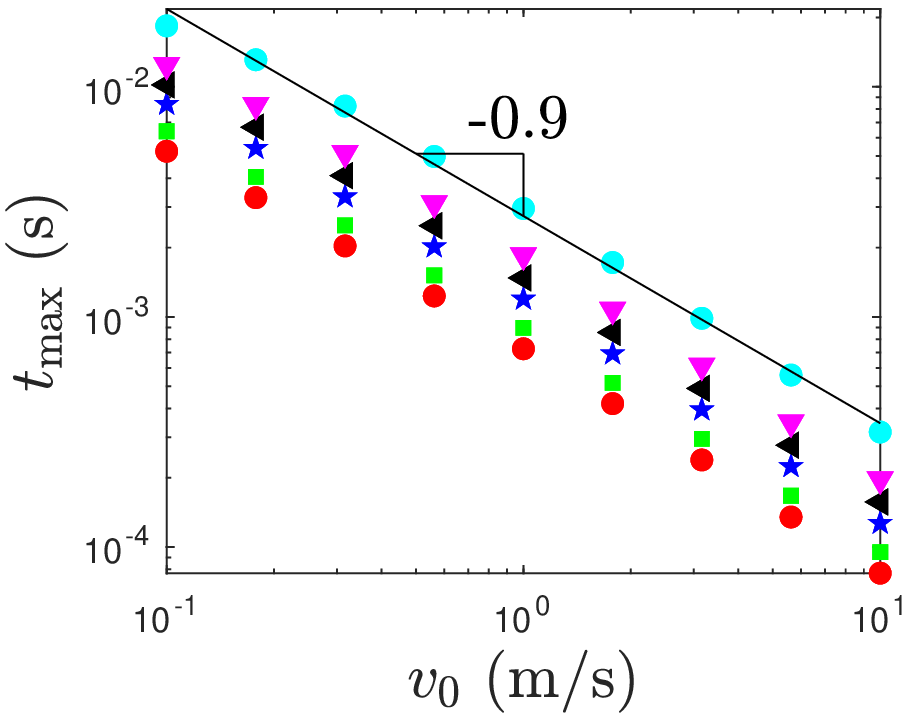} \\
    \raggedright \hspace{15mm}(c) \hspace{45mm} (d) \\ \centering
    \includegraphics[trim = 0mm 0mm 0mm 0mm, clip, width = 0.39\columnwidth]{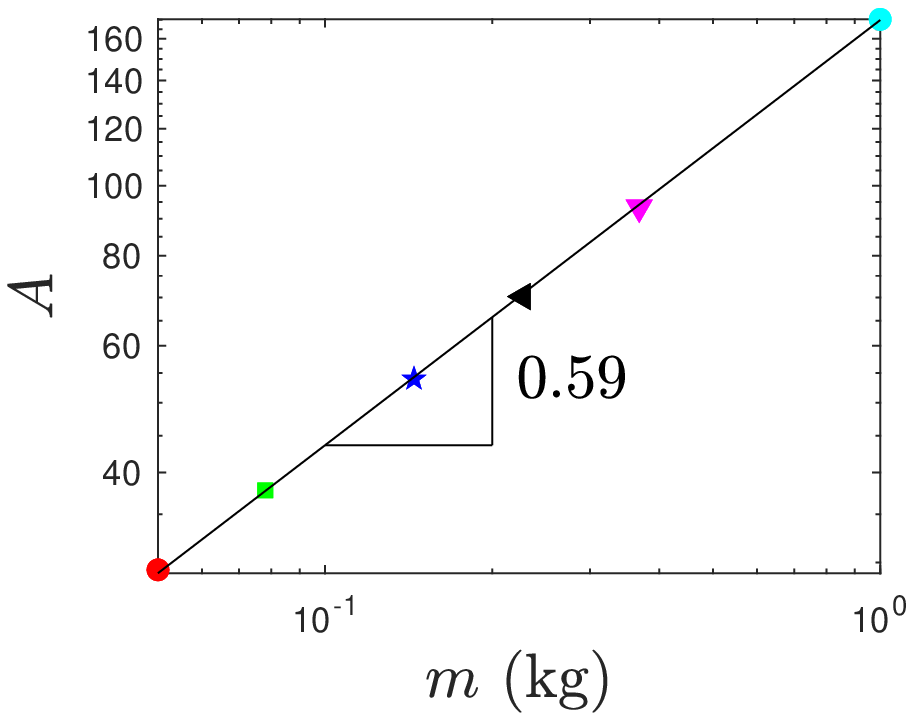}
    \includegraphics[trim = 0mm 0mm 0mm 0mm, clip, width = 0.39\columnwidth]{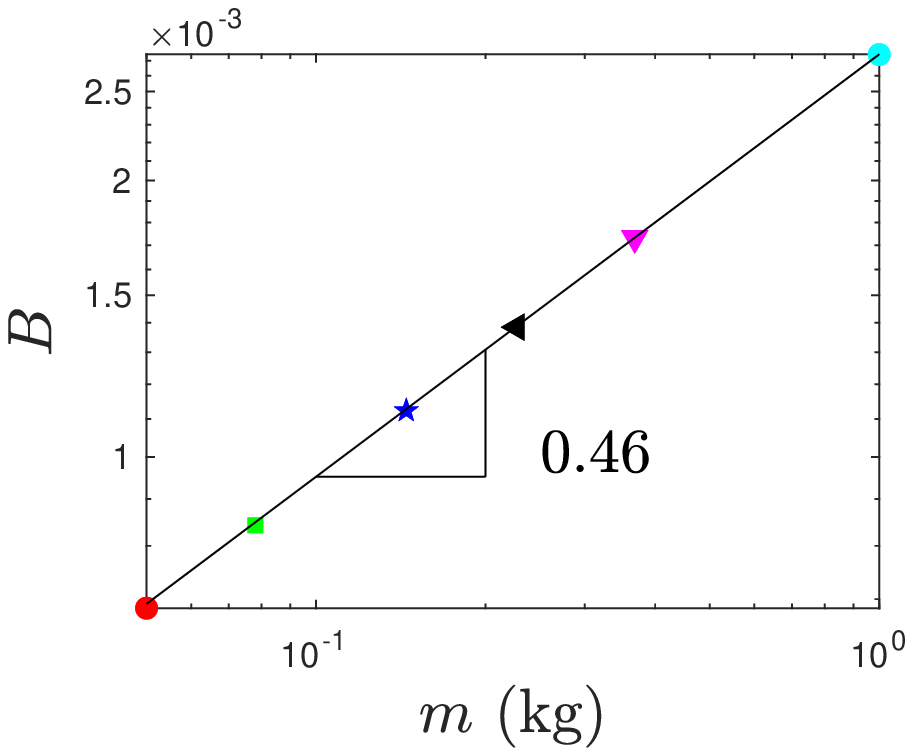}
    \caption{Mixed model numerical solutions using parameters described in the text. As in Fig.~\ref{fig:added-mass-sims}, each symbol represents a different mass.}
    \label{fig:mixed-model-sims}
\end{figure}

We also find that, if added-mass and viscous terms are both present, then the values of the exponents fall in between the predictions of each model, depending on the relative strength of each term. To illustrate this, Fig.~\ref{fig:mixed-model-sims} shows numerical solutions using Eqs.~\eqref{eqn:m_a_spread} and \eqref{eqn:F_v_spread}. We choose $C_m = 0.1$, $\rho=1630$~kg/m$^3$, $k=4$, $\Gamma = 0.7$, $C_v = 0.5$, $\eta_s = 20$ Pa$\cdot$s, and $\delta = 2$~mm, all values that are based on previous experiments or within reasonable physical bounds. This yields $F_{\rm max} \propto {v_0}^{1.61}m^{0.59}$ and $t_{\rm max} \propto {v_0}^{-0.90}m^{0.46}$.

\subsection{Summary of theoretical considerations}
\label{sec:scaling-summary}
Table~\ref{tab:summary-table} shows a summary of all the theoretical predictions from these models, assuming the forms $F_{\rm max} \propto {v_0}^\alpha m^{\zeta} D^\lambda$ and $t_{\rm max} \propto {v_0}^\beta m^{\kappa} D^\mu$. These scalings can be derived analytically only for case 2, as shown above and in~\citet{waitukaitis2014impact}, as well as case 1 in the limit of small $D$~\citep{Mukhopadyay2018}. We also solve these models numerically to confirm the analytical scalings, including the nonzero values of $D$ we use in experiments for case 1. All results from case 3 are obtained numerically.

All added mass models (case 1 and case 3 with $C_v = 0$) predict $\alpha = 2$ and $\beta = -1$. Viscous models (case 2 and case 3 with $C_v = 0$) predict $1.4<\alpha<1.63$ and $-0.67 < \beta < -0.4$. The exponents associated with $m$, $\zeta$ and $\kappa$, are similar for all models. The exponents associated with $D$, $\lambda$ and $\mu$, are zero for all added mass models, but are nonzero for the viscous model, case 2, with $\lambda = 1/2$ and $\mu = -1/2$. This is due to the fact that the size of the jammed region (and therefore the surface area experiencing viscous-like drag) scales with the intruder diameter in this model.

\begin{table}
    \centering
    \begin{tabular}{c|c|c|c|c|c|c|}
       Model & $\alpha$ & $\zeta$ & $\lambda$ & $\beta$ & $\kappa$ & $\mu$ \\
       
    \hline \hline
    Added mass (case 1) & 2 & ${2}/{3}$ & 0 & -1 & ${1}/{3}$ & 0\\
    \hline
        Viscous (case 2) & ${3}/{2}$ & ${1}/{2}$ & ${1}/{2}$ & $-{1}/{2}$ & ${1}/{2}$ & $-{1}/{2}$ \\
    \hline 
        Hybrid (case 3): $1/3 < \Gamma < 1$, $C_v = 0$, $C_m = 0.2$ & $2$ & - & 0 & $-{1}$ & - & 0 \\    \hline
        Hybrid (case 3): $\Gamma = 1$, $C_v = 0.5$, $C_m = 0$ & 1.63 & 0.69 & 0 & -0.67 & 0.33 & 0 \\    \hline
        Hybrid (case 3): $\Gamma = 1/3$, $C_v = 0.5$, $C_m = 0$ & 1.4 & 0.4 & 0 & -0.4 & 0.6 & 0 \\    \hline
        Hybrid (case 3): $\Gamma = 0.7$, $C_v = 0.5$, $C_m = 0$ & 1.58 & 0.59 & 0 & -0.6 & 0.42 & 0 \\    \hline
        Hybrid (case 3): $\Gamma = 0.7$, $C_v = 0.5$, $C_m = 0.1$ & 1.61 & 0.59 & 0 & -0.9 & 0.46 & 0 \\    \hline
    \end{tabular}    \caption{A table summarizing predictions of the models shown in Fig.~\ref{fig:front-growth}, where $F_{\rm max} \propto {v_0}^\alpha m^{\zeta} D^\lambda$ and $t_{\rm max} \propto {v_0}^\beta m^{\kappa} D^\mu$. See Section~\ref{sec:scaling-summary} for a discussion.}
    \label{tab:summary-table}
\end{table}

\section{Experiments}
\label{sec:exps}

To compare to the scaling laws summarized in Section~\ref{sec:scaling-summary}, we perform experiments of intruders falling under gravity to impact a free surface of a suspension of food-grade cornstarch particles in tap water. {The dimensions of the cornstarch suspension were roughly 20~cm $\times$ 20~cm $\times$ 20~cm (length, width, height). Figure 2 of~\citet{waitukaitis2012impact} shows that boundary effects become dominant when the fill height is less than 10~cm, meaning that our data should also be independent of boundary effects. We explicitly verified this by performing selected experiments into a suspension with larger dimensions (30~cm $\times$ 30~cm $\times$ 30~cm), and found that our results did not depend on system size.} The density of the cornstarch was 46\% by volume. We also tested impacts with 49\% by volume and found only a very slight upward shift in the forces, in agreement with~\citet{waitukaitis2012impact}. The packing fraction of cornstarch was inferred by weighing both the water and cornstarch added and assuming a specific gravity of 1.6 for the cornstarch~\citep{han2017measuring}. Intruders of varying shapes (cylinders, spheres, and cones) and diameters $D$ were attached to threaded rods and held by an electromagnet. They were held at variable heights and then released, yielding impact speeds of up to $v_0\approx 4$~m/s. The mass $m$ of the intruder was varied by adding additional weights on the rod. The impacts were recorded by high-speed video using a Phantom V711 at frame rates between 175,000 and 230,000 frames per second. A ball was attached to the threaded rod and tracked using MATLAB, yielding the position of the intruder at each frame. Discrete differentiation and a lowpass filter~\citep{Clark2012} were used to obtain the velocity and acceleration. An accelerometer with sample rate of 5000~Hz (Sparkfun ADXL377) was connected to the rod, showing good agreement with the acceleration obtained from video tracking. The accelerometer data had better time resolution, since two lowpass filters were applied to the video acceleration data. Therefore, all velocity data shown is from video tracking and all acceleration data is from the accelerometer.

\subsection{Experimental Results}

\begin{figure}
    \raggedright  (a) \hspace{65mm} (b) \\ \centering
    \includegraphics[width = 0.49\columnwidth]{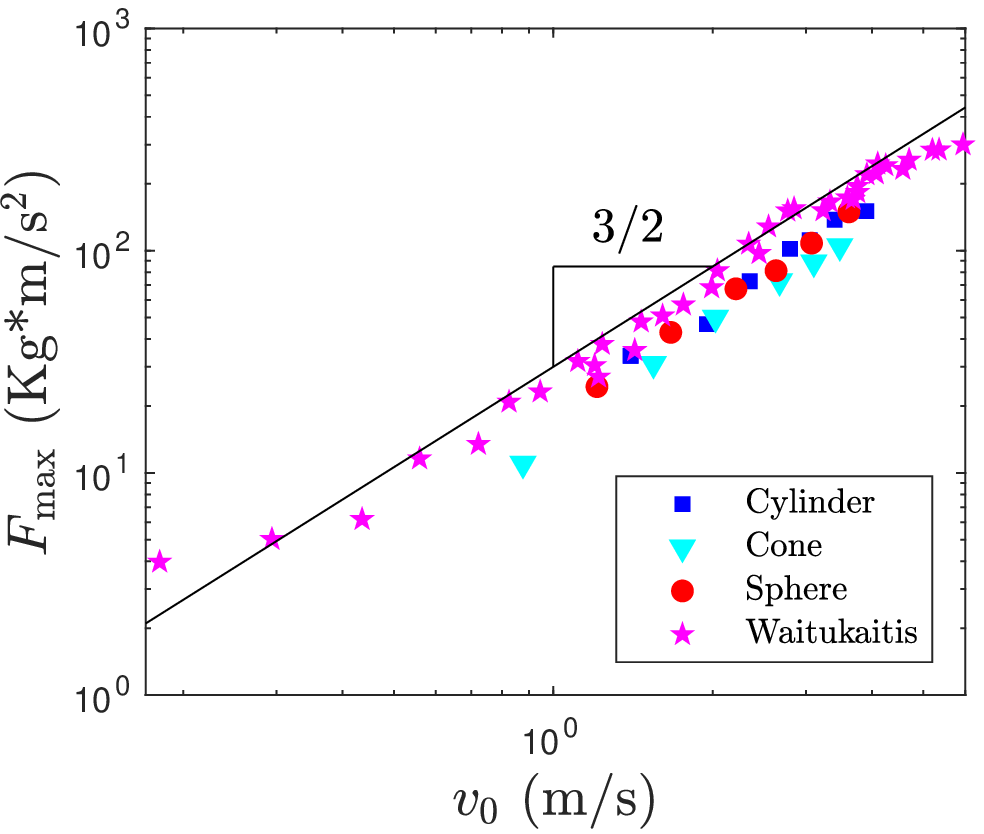}
    \includegraphics[width = 0.49\columnwidth]{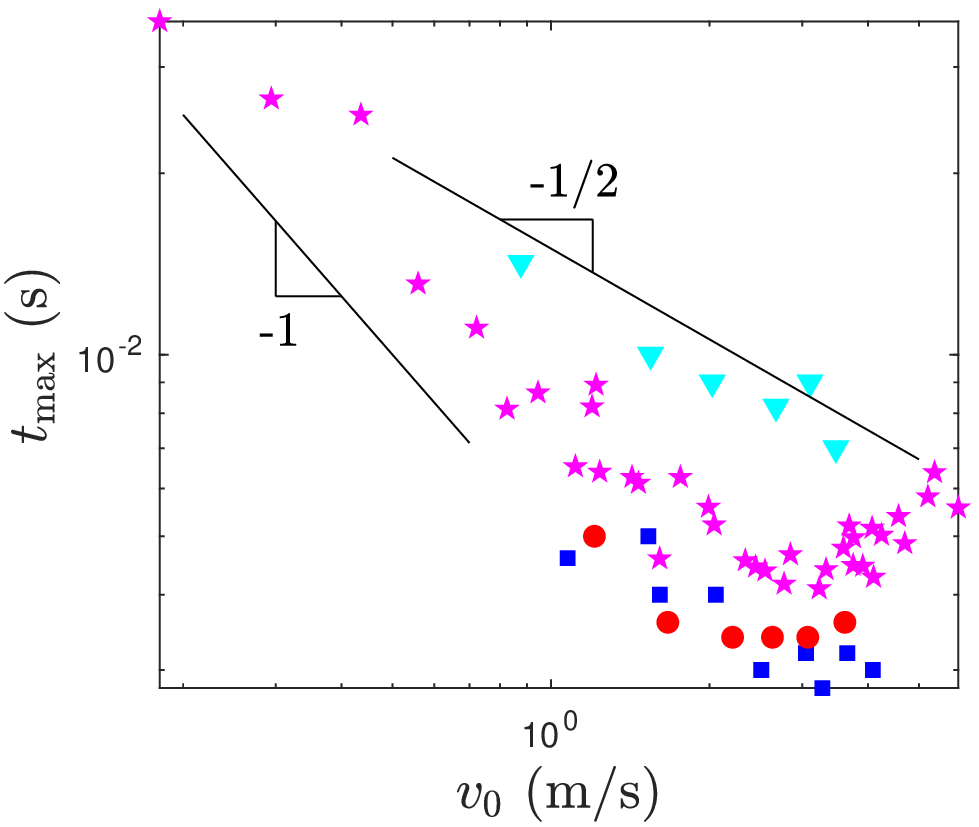}
    \caption{The magenta stars are data from Waitukaitis and Jaeger~\citep{waitukaitis2012impact} for impacts using a cylindrical intruder with $D = 1.86$~cm and $m=368$~g. The other three data sets come from our experiments, using cylindrical ($D = 3$~cm, $m = 189$~g), conical ($\theta = 70$ degrees, $D = 3$~cm, $m = 195$~g), and spherical ($D=3$~cm, $m = 199$~g) intruders. All four data sets follow power laws consistent with $F_{\rm max} = A {v_0}^{3/2}$ and $t_{\rm max}=B{v_0}^{-1/2}$; these are representative of all experiments. {We also show $t_{\rm max}\propto {v_0}^{-1}$, which is more consistent with low velocity ($v_0 < 0.5$) data from Waitukaitis.}}
    \label{fig:F-and-t-w-Wait}
\end{figure}

Figure~\ref{fig:F-and-t-w-Wait} shows $F_{\rm max} = -ma_{\rm max}$ and $t_{\rm max}$ plotted as a function of $v_0$ for four representative experiments of cylinders impacting cornstarch suspensions: three from our experiments (one cylinder, one sphere, and one cone) as well as the experimental data from~\citet{waitukaitis2012impact}. These quantities appear to scale with $v_0$ according to power law relations, Eqs.~\eqref{eq:F_max_A} and \eqref{eq:t_max_B}. Comparison with the fit line that is shown strongly suggests that $\alpha \approx 1.5$; linear fits to the data from our experiments confirm this, returning $1.3<\alpha<1.6$ for all intruders we study. This is consistent with the range predicted by viscous models discussed in Sec.~\ref{sec:theory}. The data for $t_{\rm max}$ are more scattered, making clear determination of $\beta$ more difficult. Additionally $v_0>3$~m/s, the $t_{\rm max}$ data from all intruders appear to flatten out and even curve upward slightly, which is not predicted by any of the theoretical models. However, best fits for impact velocities $0.5<v_0<3$~m/s give $\beta \approx -0.5$. These values, $\alpha = 1.5$ and $\beta = -0.5$, agree with the viscous model well, but do not agree with the predictions of the added-mass models discussed above, $\alpha = 2$ and $\beta = -1$. This strongly suggests that viscous terms at the boundary of the dynamically jammed region play an important, and likely dominant, role in the deceleration of the intruder.

\begin{figure}
    \centering
    \raggedright \hspace{15mm} (a) \hspace{45mm} (b) \\ 
    \centering
    \includegraphics[width = 0.39\columnwidth]{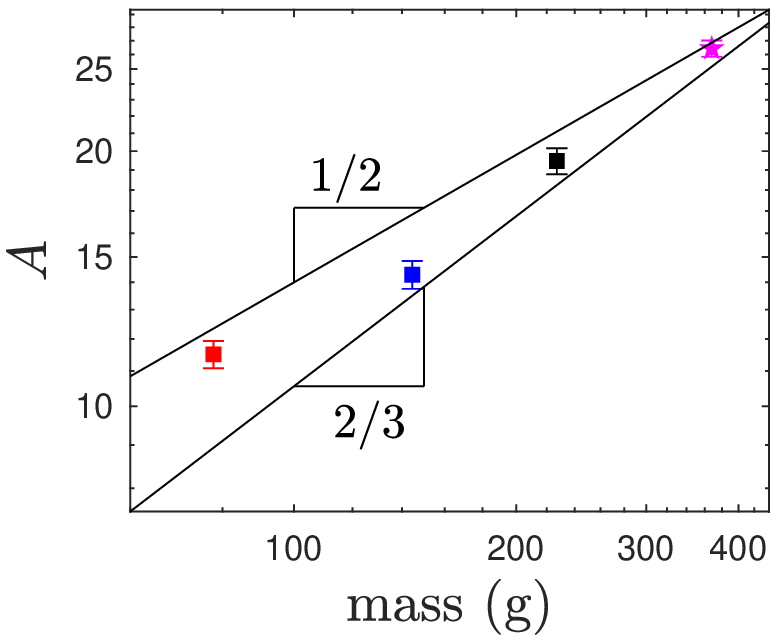}
    \includegraphics[width = 0.39\columnwidth]{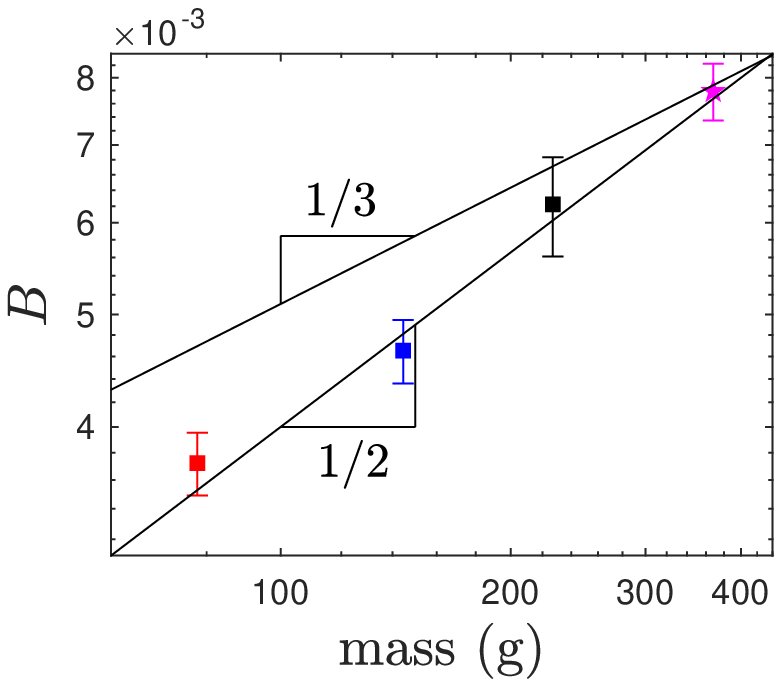}
    \\
    \raggedright \hspace{15mm} (c) \hspace{45mm} (d) \\ 
    \centering
    \includegraphics[width = 0.39\columnwidth]{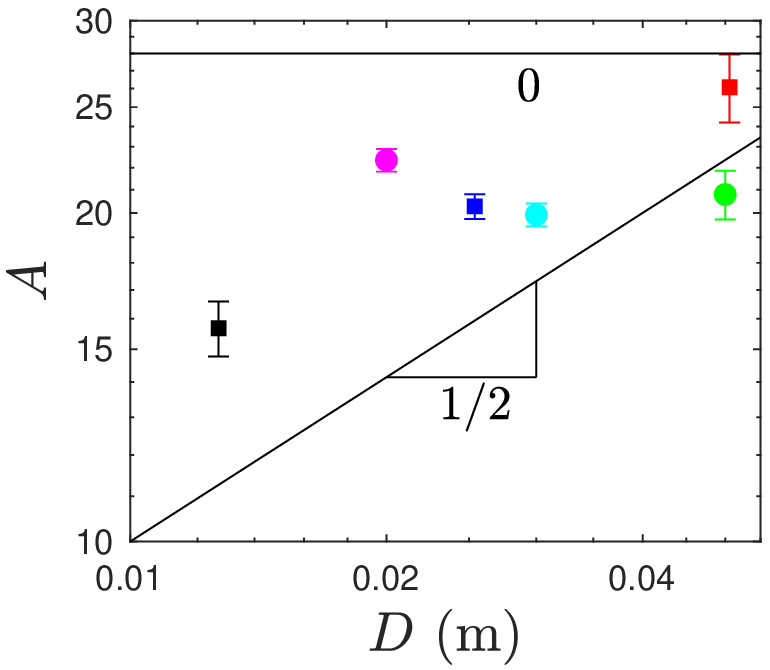}
    \includegraphics[width = 0.39\columnwidth]{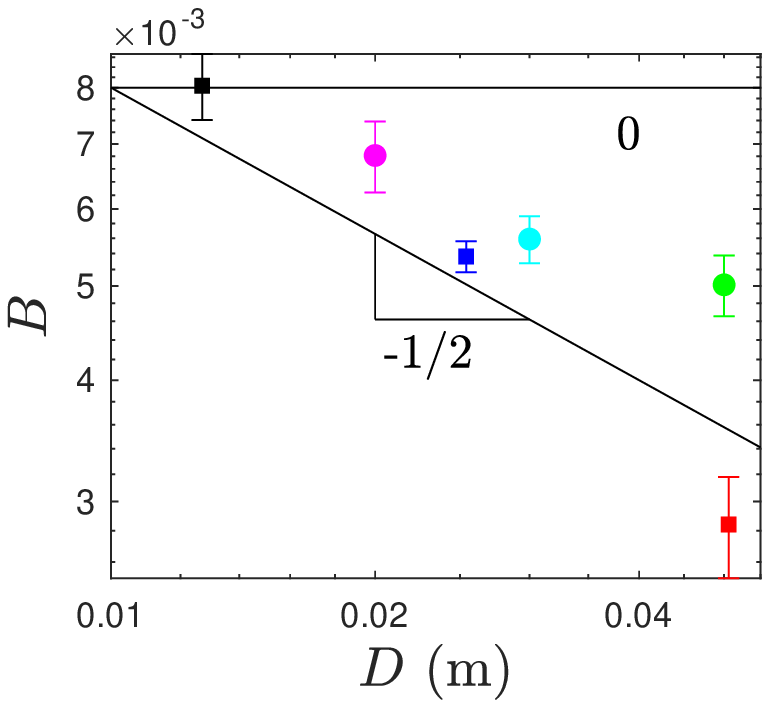}
    \\
    \raggedright \hspace{15mm}  (e) \hspace{45mm} (f) \\ 
    \centering
    \includegraphics[clip,width = 0.39\columnwidth]{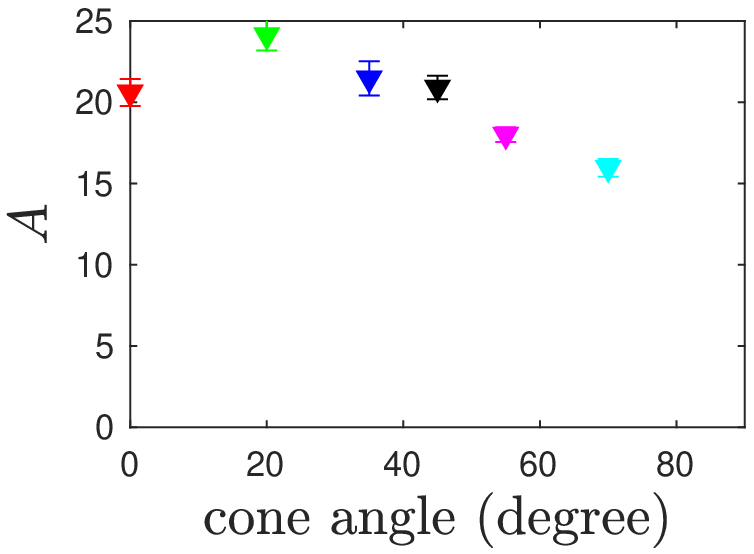} 
    \includegraphics[clip,width = 0.39\columnwidth]{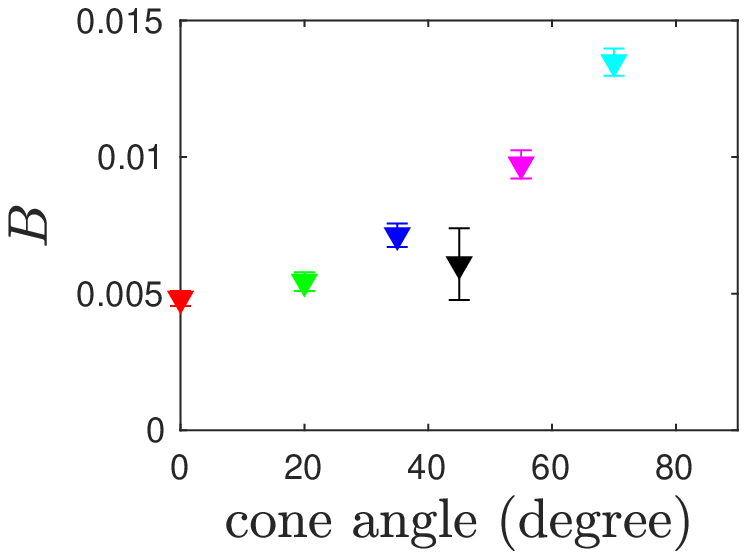} 
 
     \caption{Panels (a) and (b) show $A$ and $B$ versus $m$ for four cylindrical intruders with $D = 2.5$~cm (for the three square symbols) and $D = 1.86$~cm (for the magenta stars, from \citet{waitukaitis2012impact}). Panels (c) and (d) show $A$ and $B$ versus $D$ for experimental data on cylindrical (square symbols) and spherical intruders with $m=189$~g and $m=199$~g, respectively. Panels (e) and (f) show $A$ and $B$ for conical intruders with $m=195$~g and $D = 2.5$~cm as a function of cone angle $\theta$. Error bars represent the standard error of the mean (see text).}
     \label{fig:A-B-vs-params}
 \end{figure}

To further examine the consistency of these models with the experimental data, Fig.~\ref{fig:A-B-vs-params} shows how the prefactors $A$ and $B$ scale with $m$, $D$, and cone angle $\theta$. We measure $A$ and $B$ as the mean of $F_{\rm max}/{v_0}^{\alpha}$ and $t_{\rm max}/{v_0}^{\beta}$, {with $\alpha = 3/2$ and $\beta = -1/2$. Given the scatter in the $t_{\rm max}$ data, we also measure $B$ using other values of $\beta$ ranging from -0.4 to -1, and the results shown in Fig.~\ref{fig:A-B-vs-params} are insensitive to our choice of $\beta$. Error bars  represent the standard error on the mean; e.g., $A$ is measured as the mean of $F_{\rm max}/{v_0}^{\alpha}$, and the size of the error bar is given by the square root of these data divided by the number of data points.} Figure~\ref{fig:A-B-vs-params}(a) and (b) show $A$ and $B$ versus $m$ for three cylindrical intruders with the same diameter $D=25$~mm but with varied mass, $m \approx 80$, 150, and 230~g. Power-law fit lines are shown in black for the predictions of the original added-mass model, $A \propto m^{2/3}$ and $B \propto m^{1/3}$, and the viscous model involving a quasi-1D cylindrical dynamically jammed region, $A \propto m^{1/2}$ and $B \propto m^{1/2}$. The data appear more consistent with the viscous model predictions, especially for $B$. However, we note that the details of the shape of the added mass, as well as how dramatically the propagating front slows down as it moves, can cause these exponents to vary somewhat.

\subsection{Intruder size scaling}
Figure~\ref{fig:A-B-vs-params}(c) and (d) show data from cylindrical and spherical intruders of similar $m$ but varying $D$. The cylinders have $m \approx 190$~g and $D=12.5$, 25, and 50~mm, and the spheres have $m \approx 200$~g and $D=20$, 30 and 50~mm. The added-mass model predicts that, for these sizes and weights, there is very little dependence of $A$ or $B$ on $D$, i.e., $A \propto D^{0}$ and $B \propto D^{0}$. The first viscous model predicts $A \propto D^{1/2}$ and $B \propto D^{-1/2}$. Overall, the experimental data show that $A$ increases with $D$ and $B$ decreases with $D$, which is inconsistent with the added-mass model. We note that increase is clearer for the cylindrical intruders (square symbols) than for the spherical intruders (circular symbols). The cylindrical intruders appear to follow the predictions of the first viscous model, $A \propto D^{1/2}$ and $B \propto D^{-1/2}$. For spherical intruders, the hybrid model from Sec.~\ref{sec:hybrid-scaling} may be more relevant, where the impact is more point-like instead of a circular surface that makes simultaneous contact with the fluid. These models still predicted $\alpha \approx 1.5$ and $\beta \approx -0.5$, but they had no $D$ dependence.

\subsection{Cone shaped intruders}
Intruder shape affects $A$ and $B$ somewhat, as can be observed from dynamics of cone-shaped intruders. Figure~\ref{fig:A-B-vs-params}(e) and (f) show conical intruder data whose mass and diameter are constant, $m\approx 195$~g and diameter $D=30$~mm, but with varied angles $\theta=0^\circ$, 20$^\circ$, 30$^\circ$, 45$^\circ$, 55$^\circ$, and 70$^\circ$. Here, $\theta = 0^\circ$ corresponds to a flat cylinder and $\theta = 90$ is the maximum possible value. We observe $F_{\rm max}\propto {v_0}^{1.5}$ and $t_{\rm max}\propto {v_0}^{-0.5}$ for all cone angles, suggesting that viscous-like forces are again dominant. However, $A$ decreases with increasing $\theta$, while $B$ increases with increasing $\theta$. One hypothesis for this behavior is that larger $\theta$ corresponds to a smaller contact area, equivalent to smaller $D$. 
Another explanation could come from the fact that increasing $\theta$ means that the dynamically jammed region transitions from being generated primarily by normal compression (for $\theta = 0$) to being generated primarily through shear jamming. As shown by~\citet{Han2018shearfronts}, the value of $k$ is smaller for fronts created by shear jamming. Our data is inconclusive on this question, except for the fact that we consistently find $\alpha \approx 1.5$ and $\beta \approx -0.5$ for all values of $\theta$, which is consistent with the class of viscous models discussed in Sec.~\ref{sec:theory}.

\subsection{Relaxation after peak deceleration} \label{sec:relax}
\begin{figure}
    \hspace{5mm} \raggedright  (a) \hspace{60mm} (b) \\
    \centering
    \includegraphics[clip,width = 0.49 \textwidth]{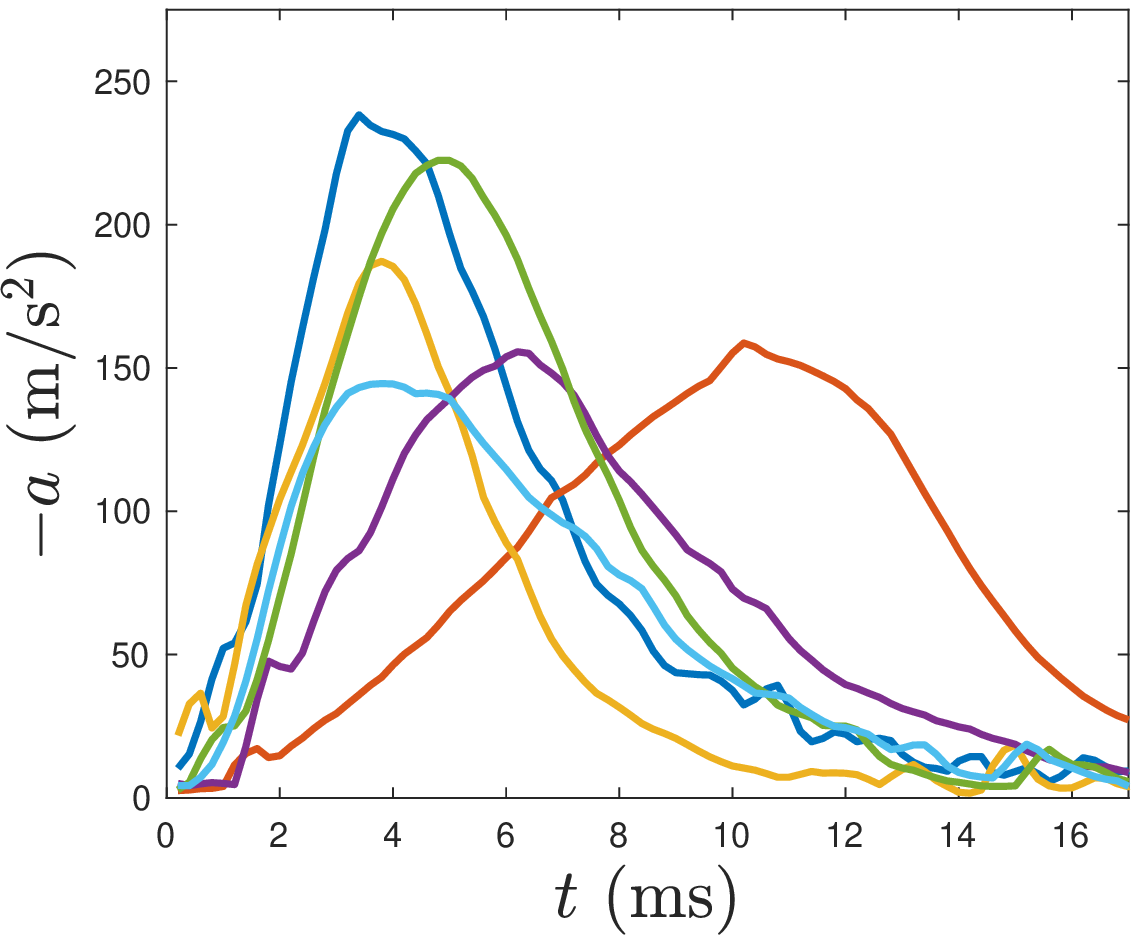}
    \includegraphics[clip,width = 0.49\textwidth]{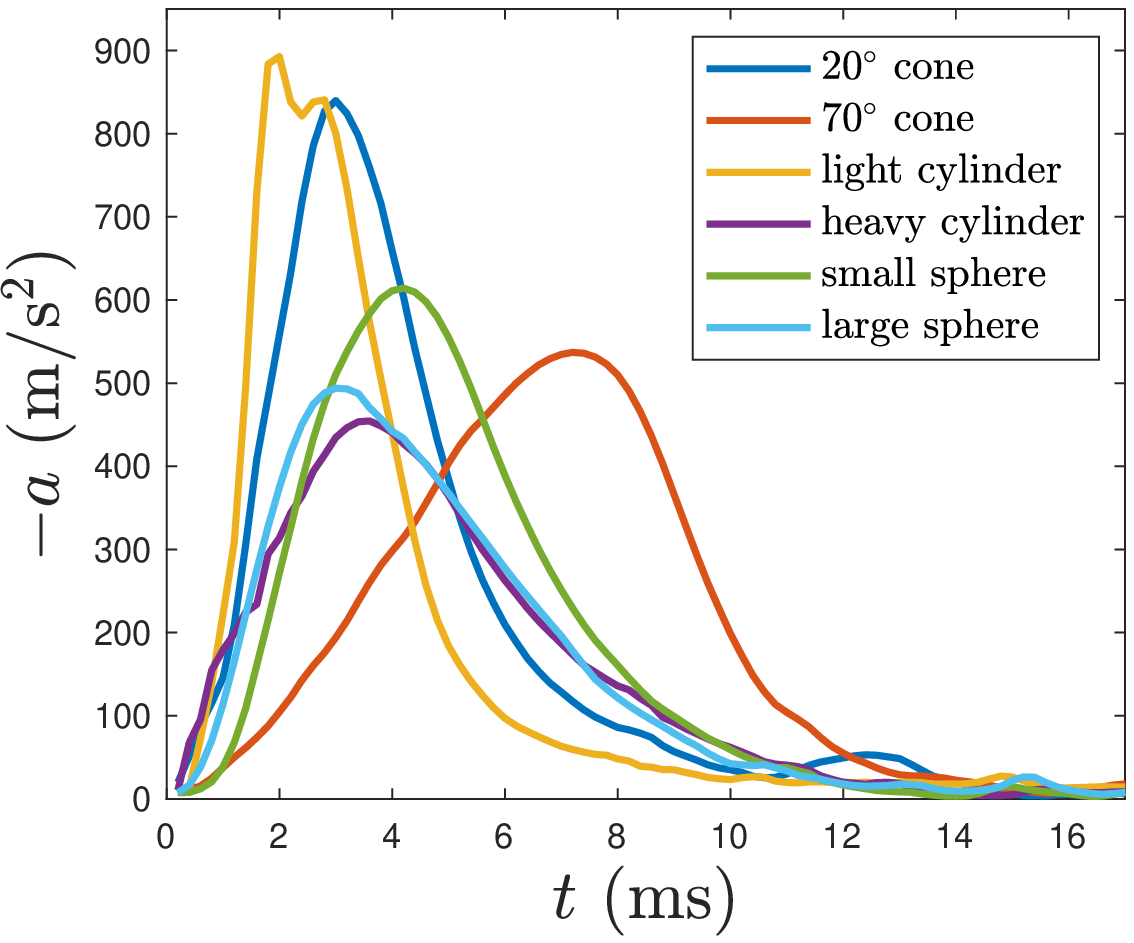} \\
    \raggedright \hspace{5mm} (c) \hspace{60mm} (d) \\
     \centering
     \includegraphics[clip,width = 0.49\columnwidth]{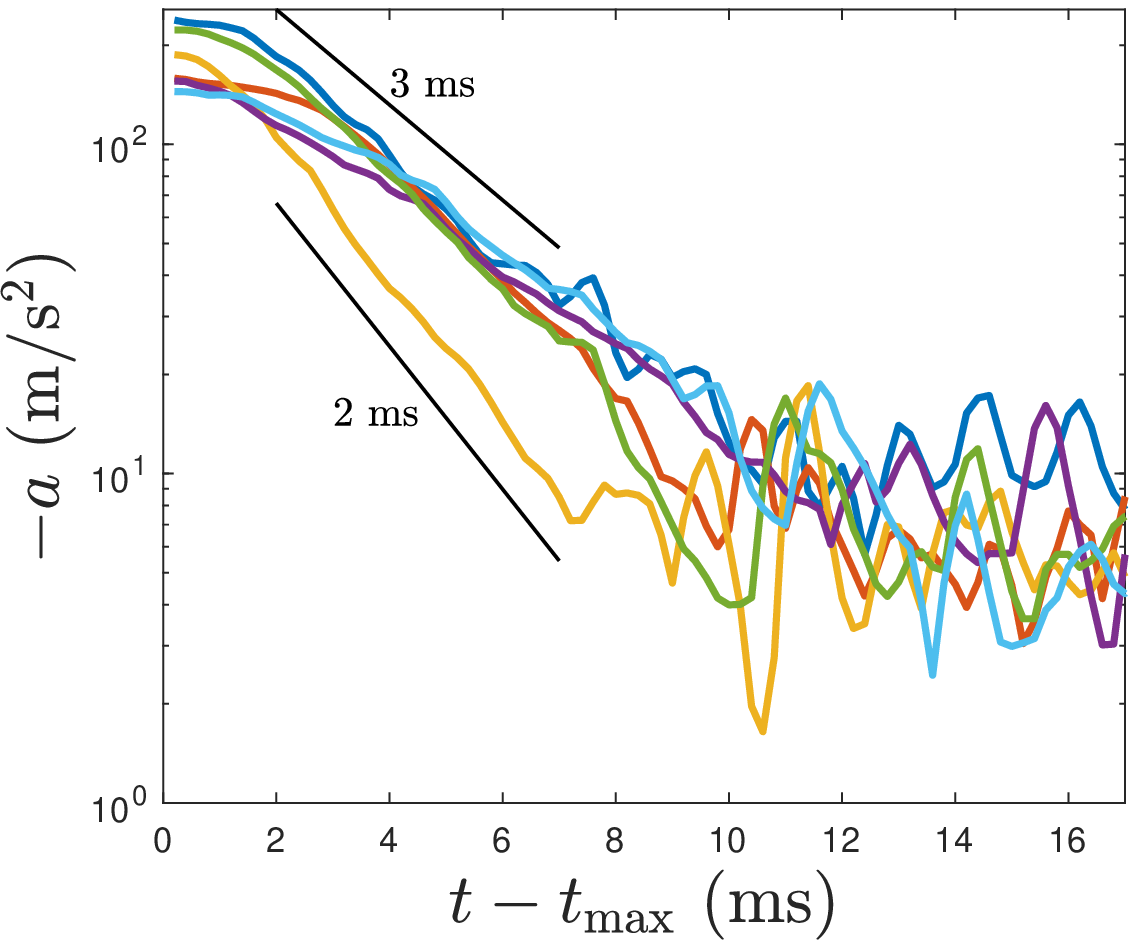}
     \includegraphics[clip,width = 0.49\columnwidth]{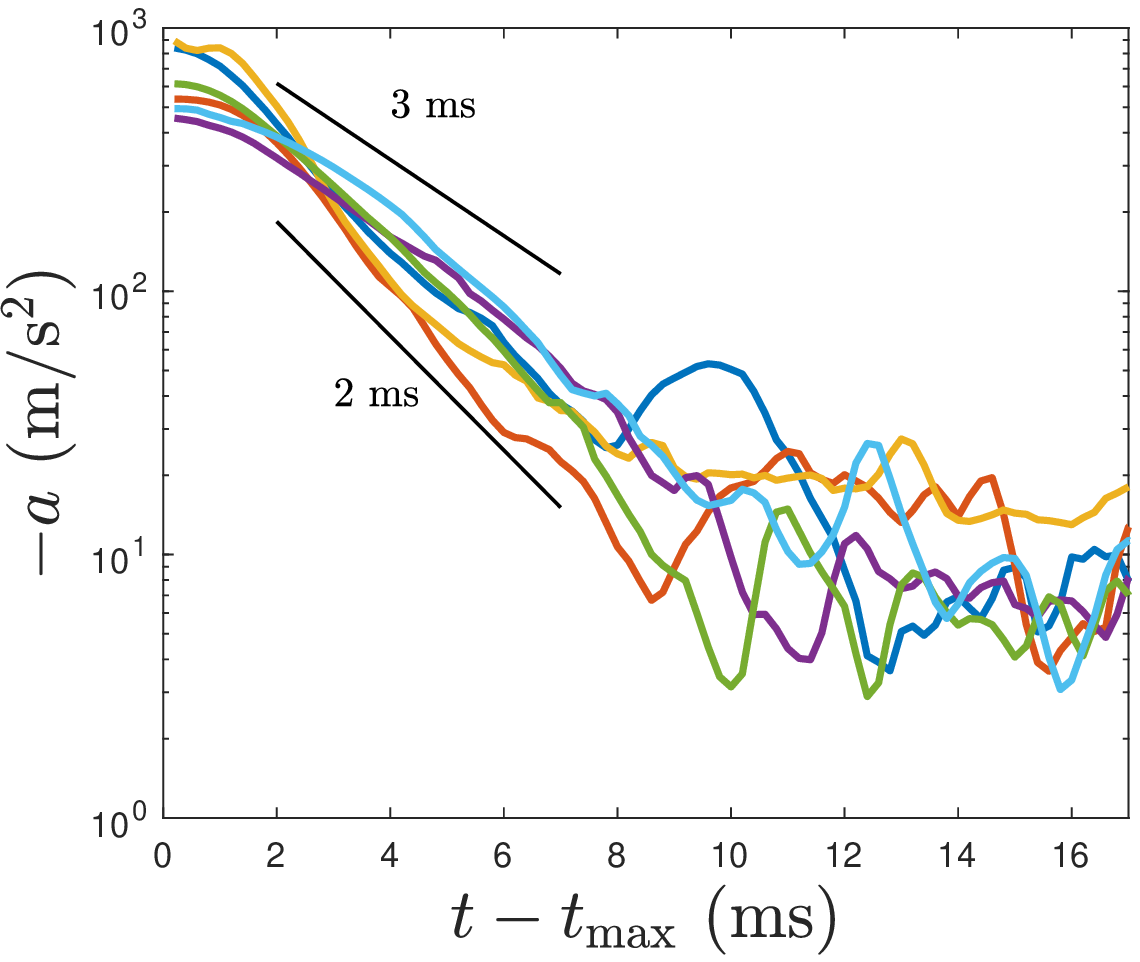}
     \caption{Acceleration curves and post-peak acceleration curves for a few representative examples of experimental data sets with (a,c) $1<v_0<1.5$~m/s and (b,d) $3<v_0<3.5$~m/s. On (c) and (d), we indicate exponential decay with time scales of 2 and 3 ms with solid lines in both panels for reference. The cones have $m=195$~g and $D = 2.5$~cm; the light and heavy cylinders have $D = 2.5$~cm with $m = 77.8$~g and $m = 227$~g, respectively; and the small and large spheres have $m=199$~g and $D=2$~cm and $D=5$~cm, respectively.}
     \label{fig:decay}
\end{figure} 
Finally, we consider the intruder dynamics after the peak deceleration, which also provides information on the microscopic physics of suspension dynamics. While the dynamics of impact before peak deceleration are highly sensitive to various experimental control parameters, the post-peak dynamics are not. This is shown in Fig.~\ref{fig:decay}. Figure~\ref{fig:decay}(a) and (b) show impacts with $1<v_0<1.5$~m/s and $3<v_0<3.5$~m/s, respectively, both with a wide variety of intruder properties. $t_{\rm max}$ varies dramatically with $v_0$ and intruder properties, which has been the subject of our analysis so far. However, Fig.~\ref{fig:decay}(c) and (d) show that forces decay quasi-exponentially with a time scale between 2 and 3 ms; this behavior is largely independent of speed and intruder properties. This implies that, e.g., the time scale $\tau = \sqrt{C_v k \eta_s v_0/2m}$ in Eqs.~\eqref{eqn:v-viscous} and \eqref{eqn:a-viscous} might capture the buildup to peak force but not the decay. This suggests that the relaxation dynamics are dominated by the microscopic material composition in a way that is not sensitive to the intruder. Thus, these dynamics appear to lie outside the description of either the added mass or viscous models. Our observations are consistent with~\citet{peters2014quasi}, who found that changing the viscosity of the suspending fluid affected the dynamics of the relaxation of the jammed front but not of its growth.
 
{The fact that fluid viscosity slows down the relaxation dynamics suggests that Darcy-Reynolds theory, demonstrated by~\citet{jerome2016}, may play a crucial role in holding the solidified region together. The system studied by \citet{jerome2016} involves much larger, inorganic glass beads with diameter of roughly 100~$\mu$m or larger, whereas (organic) cornstarch particles are roughly 1 to 10~$\mu$m. The grains were gravitationally loaded into a settled configuration--that is, the initial state of the system involves all particles making solid-solid contact--as opposed to the cornstarch particles, which are stirred between impacts and remain suspended in the fluid at relatively low volume fractions. Then, if the grains dilate during impact, there is a strong solidification that is stricly driven by a combination of Reynolds dilatancy (the packing expands) and Darcy flow (high resistance of the fluid being sucked into the expanding pores between grains); if the grains compact, there is no solidification. Our system begins at a low volume fraction and likely does not compact significantly~\citep{han2016high}. Thus, it is unlikely that the Darcy-Reynolds theory plays a direct, dominant role in the present experiments, but it may play a crucial, secondary role. Relaxation of the solidified region occurs as particles rearrange and lose contact with each other, which requires fluid flow through the pore structure. Thus, Darcy flow in particular may be responsible for holding the solidified region together and may also control the relaxation dynamics shown in Fig.~\ref{fig:decay}.}

\section{Conclusion}
\label{sec:conclusion}

Here we have theoretically and experimentally studied the problem of impact of an intruder into a shear-thickening dense suspension. In agreement with previous authors such as~\citet{Mukhopadyay2018}, we find that the added-mass model~\citep{waitukaitis2012impact}, which has been the dominant model used to explain the dynamics of impacts into shear-thickening dense suspensions, predicts $F_{\rm max} \propto {v_0}^2$ and  $t_{\rm max} \propto {v_0}^{-1}$. In contrast, the experimental data show $F_{\rm max} = A {v_0}^{1.5}$ and  $t_{\rm max} = B {v_0}^{-0.5}$. These exponents are consistent with a class of models where the dominant force is not added mass but viscous-like forces at the boundary of the jammed suspension. 

We have also studied how the prefactors $A$ and $B$ depend on intruder mass, size, and shape. These results are either consistent with both added-mass and viscous models (e.g., in the case of varying intruder mass) or more consistent with viscous models (e.g., in the case of cylindrical intruders with varying diameter). Our results suggest that the added-mass model should be revised to include large, viscous-like terms at the boundary, since these forces may play a dominant role. These results do not change certain aspects of the underlying physical picture for impact into dense suspensions: a solid-like region grows outward from the point of impact and dominates the intruder dynamics. If {large, viscous-like forces (corresponding to the large, nearly constant viscosity observed in the shear-thickening regime of dense suspensions) are dominant}, this also has the advantage of conceptually unifying impact with steady-state rheology descriptions like DST.

We emphasize that the viscous models we analyze here are not applicable to simple, high-viscosity liquids. The models we discuss are predicated on the existence of a growing solid-like region and a large viscosity that is ``turned on'' only in regions of very high strain rate. These features would not exist in the case of a simple liquid with very large viscosity, as the viscosity would be constant everywhere (independent of local shear rate) and there would be no solid-like region.

As mentioned in Sec.~\ref{sec:intro}, we note that Fig.~6 of \citet{peters2014quasi} shows that added mass is sufficient to explain the forces measured by an external sensor for velocity controlled impact into a 2D layer of dense suspension. However, in a 2D experiment, the viscous-like forces that we propose would act over a 1D boundary between the jammed region and the uncompressed suspension; in a 3D experiment, the surface area of the 3D jammed solid is much bigger, leading to significantly larger viscous forces. Thus, their findings (that added mass was sufficient to explain the resisting force on the intruding object in a 2D situation) are consistent with the results we have shown for 3D impacts.

Our results are limited to the case where the jammed region does not span the system, which leads to a ``bounce'' as shown in Fig.~2 of~\citet{waitukaitis2012impact}. The mechanical response of a system-spanning dynamically jammed region was analyzed by~\citet{allen2018} and \citet{Maharjan2018} in the case of a much smaller container with an intruder driven at constant speed. Thus, we expect the impact process to be significantly different in the case of a shear-thickening suspension in a smaller container or confined to a thin layer.

Finally, we reiterate that we have made several approximations throughout our analysis, such as neglecting buoyant and gravitational forces. Thus, inclusion of these or other forces may help resolve the discrepancies between the models and experimental results. Future work is needed to better characterize the relative contribution of these forces; to better characterize the magnitude of and the length scale over which the viscous forces act; and to understand how the solidified mass relaxes back into a fluid-like state.

\acknowledgements
This article was made possible by the Office of Naval Research under Grant No. N0001419WX01519 and by the Office of Naval Research Global Visiting Scientist Program VSP 19-7-001. We thank Scott Waitukaitis for sharing his data and for helpful discussions on his PhD thesis. Declaration of Interests: the authors report no conflict of interest.

\bibliographystyle{jfm}
\bibliography{references}
\end{document}